\documentclass[a4paper,fleqn]{cas-sc}

\usepackage[authoryear]{natbib}
\usepackage{color}

\begin{document}

\let\WriteBookmarks\relax
\def\floatpagepagefraction{1}
\def\textpagefraction{.001}

\shorttitle{Streams of comet 21P on Mercury, Venus, and Mars}    
\shortauthors{D. Tomko \& L. Neslu\v{s}an}  


\title[mode = title]{Prediction of the collisions of meteoroids
       originating in comet 21P/Giacobini-Zinner with the Mercury,
       Venus, and Mars}



\author[1]{D. Tomko}

\address[1]{Astronomical Institute, Slovak Academy of Science,
 05960 Tatransk\'{a} Lomnica, Slovakia}



\author[1]{L. Neslu\v{s}an}[orcid=0000-0001-9758-1144]

\ead{ne@ta3.sk}



\cortext[cor1]{Corresponding author: L. Neslu\v{s}an}


\begin{abstract}
After the prediction of meteor showers in the Earth's atmosphere caused
by the particles originating in the nucleus of comet
21P/Giacobini-Zinner, we went on with the prediction of showers on the
other three terrestrial planets. Based on our modeling of theoretical
stream of the parent comet, we predicted several related meteorite (on
Mercury) or meteor (on Venus and Mars) showers. There occurred the
filaments, in the stream, with the particles coming to each planet from
a similar direction. We found that this is a consequence of the specific
distribution of argument of perihelion (peaked close to the value of
$180^{\circ}$) and longitude of ascending node of the stream, and that
the particles collide with each planet in an arc of their orbits being
close to perihelion.
\end{abstract}




\maketitle

\section{Introduction}

   A number of meteoroid streams move in the interplanetary space. When
a planet crosses the spatial corridor in which a stream moves, the
meteoroid particles collide either with its atmosphere or surface. A
meteor shower can then be detected. We know a lot of meteor showers that
can be observed in the atmosphere of our planet, the Earth.

   However, the collisions of meteoroids also happen with the other
planets. At the present, there are no observational stations to detect
the meteors (or meteorites) on the other planets. However, we can
predict an existence of meteor showers if other planet passes,
periodically, through a corridor of a stream, which is colliding with
the Earth. Or, we can follow the dynamical evolution of stream
originating in a particular cometary nucleus or asteroid, when we know
or assume that the object is a source of meteoroid particles.

   In the past, the meteoroid streams colliding with the other
terrestrial planets were studied by several authors \citep{Beech1998,
Christou_etal2007, McAuliffe_Christou2005, Neslusan2005,
Withers_etal2007, Espley_etal2010, Christou_etal2012, Dmitriev_etal2013,
Withers_etal2013, Christou_etal2015a, Christou_etal2015b,
Fries_etal2015, Fries_etal2016, Kuznetsova_etal2018}. The comprehensive
review of meteor showers, which could probably be detected on Venus and
Mars, was given by \citet{Christou2010}. Recently,
\citet{Christou_etal2019} published the review chapter about the
``extra-terrestrial'' meteors.

   In this work, we deal with the meteoroids originating in the periodic
comet 21P/Giacobini-Zinner in course to predict their collisions with
the terrestrial planets Mercury, Venus, and Mars. In our earlier work
\citep[][Paper I, hereafter]{Neslusan_Tomko2023}, we already modeled the
stream of this comet and predicted the associated meteor showers
observable in the Earth's atmosphere.

\section{Method and models created}

   The method how the models of 21P's meteoroid stream were created was
described in Paper I. These models can be used to predict the meteor
showers observable at any terrestrial planet; in the atmosphere of
Venus and Mars as well as the showers of meteorites impacting the
surface of Mercury.

   We briefly remind the method of model creation. In the first step,
the nominal orbit of the parent comet, 21P, was integrated in time
backward for an arbitrarily selected period, $t_{ev}$. In more detail,
the integration was terminated exactly in the time of the comet's
perihelion passage that was nearest to time $t_{ev}$ before the present.
In the perihelion, we assumed an ejection of 10,000 test particles from
the comet's nucleus. The particles were ejected randomly to all
directions with the same ejection speed equal to one thousandth
heliocentric speed of the comet in its perihelion. The aim of this
stream modeling was filling in the orbital phase space of the assumed
stream with the particles; we did not intend to reproduce a real
ejection of meteoroids from the surface of their parent body. In a
further step, we followed, via a numerical integration, the dynamical
evolution of the created set of the test particles up to the present,
i.e. for the period of $t_{ev}$. Hence, $t_{ev}$ is referred to as an
evolutionary time.

   In the forward integration, we assumed the Poynting-Robertson (P-R)
effect influencing the motion of the particles \citep{Klacka2014}. The
strength of this effect is characterized with a dimensionless parameter
$\beta$. This parameter and the evolutionary time are supposed to be the
free parameters in our modeling. In accord with Paper I, the models for
all combinations of the values $t_{ev} = 0.5$, 1, 2, and $4\,$kyr and
$\beta = 10^{-11}$, $10^{-4}$, 0.001, 0.003, and 0.005 were considered.
The results yielded by these models are described in the following
section. The numerical integration was performed by using the integrator
Gauss-Radau (RA15) \citep{Everhart1985} within the MERCURY software
package \citep{Chambers1999, Chambers_Murison2000}.

\section{Results}

   In all models, the meteoroid stream splits to several filaments,
the particles of which can be detected as individual meteor or
meteorite showers on given planet. The mean planet-centric parameters
and mean orbital elements of these filaments are given in
Tables~\ref{TAB1MERC}$-$\ref{TAB2MARS}. These parameters were
determined by averaging the parameters of individual particles with
the minimum-orbit intersection distance (MOID), in respect to given
planet, smaller than $0.02\,$au at the end of the numerical integration
(i.e. at the present time). In tables, we denote the northern filaments
with abbreviation ``NF'' and southern filaments with ``SF''. We assigned
a number to each filament, which follows the NF or SF.

   The radiants of particles in selected models of each filament are
shown in Figs.~\ref{hammerMerc}$-$\ref{hammerMars}. In more detail, we
selected the model, in which the given filament corresponded to the
shower of category (i) (see below) and was represented by the largest
number of particles. If there is no shower of category (i), then we
selected the model of category (ii) with the largest number of
particles. In the figures, the filaments are mutually distinguished by
various colors and symbols.
   Although we describe the radiants on the sky of a planet other than
Earth, we still use the common equatorial or ecliptical coordinate
frames; it would be difficult and, likely, useless to define other,
local, coordinate systems.

   As stated by \citet{Marsden_Sekanina1971} and confirmed in our Paper
I, comet 21P has been in an erratic, rapidly changing orbit. This have
caused a fast change of orbits in its meteoroid stream. The orbits of
particles are dispersed enough and the filaments have often not well
defined orbital corridor. Due to this circumstance, it was, in some
cases, hard to decide if an accumulation of radiants or orbits is or is
not a shower.

   Thus, we divided the showers into three categories: (i) well
recognizable, (ii) dispersed but still recognizable, and (iii) extremely
dispersed. The division was rather intuitive than exact; we usually
classified a shower in the first category when the determination
errors ($\sigma$s) of its argument of perihelion, $\omega$, and
longitude of ascending node, $\Omega$, were lower than
$\sim$$10^{\circ}$. Sometimes, there was a grouping of radiants, but
$\sigma$ of $\omega$ or $\Omega$, or both exceeded $\sim$$40^{\circ}$.
Such a shower, even if really existed, could scarcely be recognized in
a database of meteors or meteorites. We classified such very dispersed
group of particles as belonging to category (iii). Finally, these
groupings were omitted in a further consideration; they are not listed
in the tables. The lines with the mean parameters of the showers of
category (ii) are given in the brackets in the tables.

\subsection{21P's meteorite showers observable on Mercury}

   Mercury is a planet without an atmosphere that could act on the
particles moving toward its surface. Therefore, all meteoroids in the
collisional course with this planet impact its surface. One can suppose
that the microscopic particles are incorporated to the surface material
of the planet. The larger particles obviously survive as the meteorites.

   Term ``meteor'' refers to a phenomena in a planetary atmosphere, i.e.
in a gaseous environment. A stream of meteoroids leads to a
corresponding meteor shower. In the case of Mercury, this term should
not be used. We will rather speak about a meteoroid stream impacting
the first Solar-System planet.

\begin{figure*}
\centerline{\includegraphics[width=7.5cm, angle=-90]{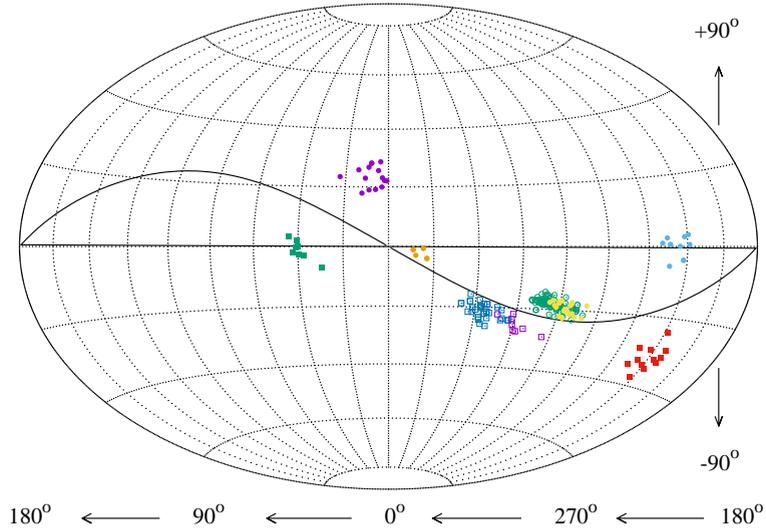}}
\caption[f2]{Positions of the radiants of theoretical particles in the
atmosphere of Mercury associated with comet 21P. The positions are
plotted in the Hammer projection of the celestial sphere. Equatorial
coordinate frame is used. The sinusoid-like curve indicates the
ecliptic. The radiants of the theoretical particles in the predicted
filaments on the northern hemisphere NF1, NF3, NF4, and NF5 are shown
with purple, turquoise, orange, and yellow full circles. Filament NF2 is
shown with green empty circles. The radiants of the theoretical
particles in the predicted filaments SF2 and SF3 on the southern
hemisphere are shown with red and green full squares and filaments SF1
and SF4 with blue and purple empty squares. Full symbols are used for
the radiants of showers of category (i) and empty symbols for category
(ii).}
\label{hammerMerc}
\end{figure*}

   Also in the case of Mercury, the 21P' stream splits into several
filaments colliding with the planet. In particular, the modeling of
meteoroid stream resulted in the prediction of five Mercury's meteoroid
filaments with radiants on the northern sky and five filaments with
radiants on the southern sky. The mean parameters characterizing these
filaments are listed in Tables~\ref{TAB1MERC} (mean planet-centric
parameters) and \ref{TAB2MERC} (mean orbital elements). The radiants of
the particles used to predict the filaments in selected models are
in Fig.~\ref{hammerMerc}.

   The radiants of filaments NF1, NF3, and NF4 (purple, turquoise, and
orange full circles, respectively) are well separated each other as seen
in Fig.~\ref{hammerMerc}). The radiants of NF2 and NF5 (empty green and
full yellow circles) are overwhelming. The particles of these two
filaments however clearly differ each other by solar longitude
($(80.0^{\circ}$$-$$101.8^{\circ}$ vs. $327.3^{\circ}$$-$$352.5^{\circ}$)
and argument of perihelion ($323.2^{\circ}$$-$$340.4^{\circ}$ vs.
$6.6^{\circ}$$-$$23.9^{\circ}$).

   Also southern filaments SF1, and SF4 (blue and purple empty squares)
have a common radiant area and can be distinguished by largely different
solar longitude and argument of perihelion. Only SF2 and SF3 (red and
green full squares) can be well distinguished among the other southern
filaments.

\subsection{21P's meteor showers observable on Venus}

\begin{figure*}
\centerline{\includegraphics[width=7.5cm, angle=-90]{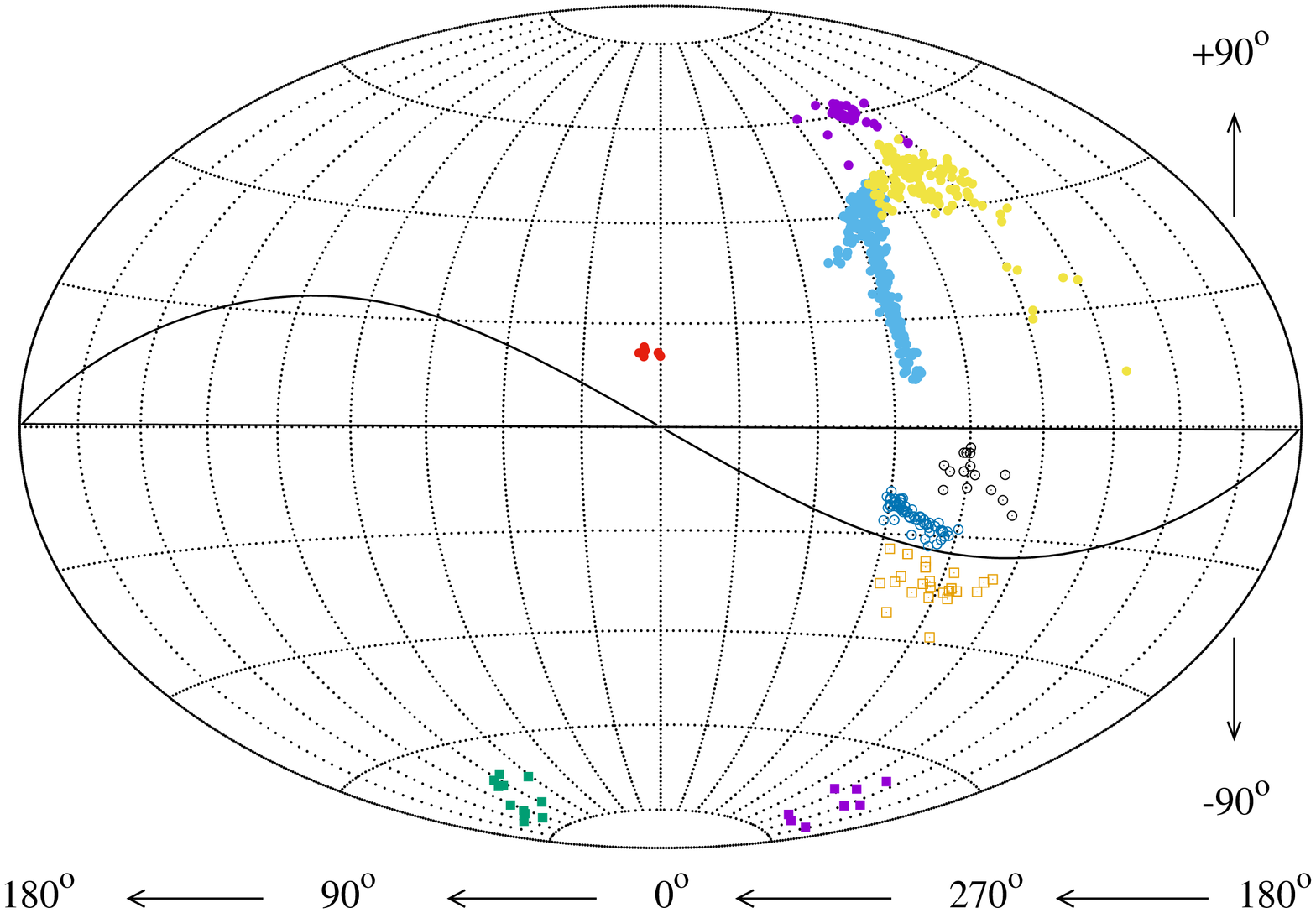}}
\caption[f2]{Positions of the radiant of theoretical particles in the
atmosphere of Venus associated with comet 21P.The positions are plotted 
in the Hammer projection of the celestial sphere. Equatorial coordinate
frame is used. The sinusoid-like curve indicates the ecliptic. The 
radiants of the theoretical particles in the predicted filaments on the
northern hemisphere NF1, NF2, NF3, and NF5 are shown with purple,
turquoise, yellow, and red full circles. Filaments NF4 and NF6 are shown
with green and black empty circles. The radiants of the theoretical
particles on the southern hemisphere in the predicted filament SF1 are
show with orange empty squares and those of filaments SF2 and SF3 with
purple and blue full squares. Full symbols are used for the radiants of
showers of category (i) and empty symbols for category (ii).}
\label{hammerVen}
\end{figure*}

   Venus moves through the corridors of six northern and three southern
filaments of stream of 21P (Fig.~\ref{hammerVen}, Tables~\ref{TAB1VEN}
and \ref{TAB2VEN}).

   There is an interesting radiant area of filament NF2 (full turquoise
circles in Fig.~\ref{hammerVen}). It is narrow, but with a large extent
in declination. This area is not separated from the radiant area of
filament NF3 (full yellow circles), but filaments NF2 and NF3 can be
well distinguished by their different mean angular distance from the
Sun, $\gamma$, in time corresponding to the mean solar longitude.
While $\gamma \geq 121^{\circ}$ in NF2, $\gamma < 86^{\circ}$ in NF3.
The solar longitude, mean radiant, and mean angular orbital elements
are also different.

   Southern filament SF1 was predicted almost in every model that we
constructed. However, the radiants of individual particles were,
sometimes, so much dispersed that we could not regard it as a shower.

   Filaments NF4, NF5, and NF6 were predicted in only two models. In
NF4 and NF6, the parameters of the individual particles are relatively
dispersed (showers of category (ii)). NF5 (red empty circles in
Fig.~\ref{hammerVen}) was predicted only with the help of a low number
of particles. However, it is a quite compact, well-defined shower.


\subsection{21P's meteor showers observable on Mars}

\begin{figure*}
\centerline{\includegraphics[width=7.5cm, angle=-90]{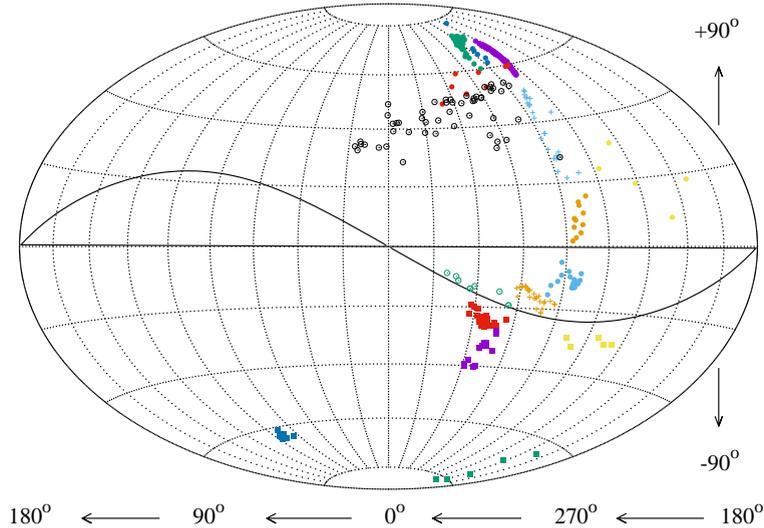}}
\caption[f2]{Positions of the radiants of theoretical particles in the
atmosphere of Mars associated with comet 21P. The positions are plotted
in the Hammer projection of the celestial sphere. Equatorial coordinate
frame is used. The sinusoid-like curve indicates the ecliptic. The
radiants of the theoretical particles in the predicted filaments on the
northern hemisphere NF1, NF2, NF5, NF6, NF7, NF10, and NF11 are shown
with purple, green, yellow, blue, red, turquoise, and orange full
circles, respectively. Filaments NF3 and NF4 are shown with turquoise
and orange crosses, and filaments NF8 and NF9 with black and green empty
circles. The radiants of the theoretical particles in the predicted
filaments on the southern hemisphere of sky SF1, SF2, SF3, SF4, and SF5
are shown with yellow, blue, red, purple, and green full squares. Full
symbols and crosses are used for the radiants of showers of category (i)
and empty symbols for category (ii).}
\label{hammerMars}
\end{figure*}

   Our modeling of 21P' stream (described in Paper I) also predicted
some showers, which could be observable in the atmosphere of Mars. Their
mean planet-centric parameters are given in Table~\ref{TAB1MARS} and
mean orbital elements in Table~\ref{TAB2MARS}. The positions of radiants
of particles in the individual filaments, as predicted by the selected
models, are shown in Fig.~\ref{hammerMars}.

   In total, the modeling predicted eleven northern and five southern
showers, observable on this planet, which originated in 21P. Some of
the predicted filaments (NF5, NF6, NF7, NF9, and NF11) are, however,
questionable and should be considered with a caution. Filament NF8
(radiants shown with the black empty circles in Fig.~\ref{hammerMars})
has also a largely dispersed radiants. We eventually consider it (shower
category (ii)) since the values of its mean perihelion distance,
eccentricity, argument of perihelion, longitude of ascending node, and
geocentric velocity are predicted very similar in all five models that
yielded this filament.

   In Fig.~\ref{hammerMars}, one can see a complex of 21P' showers
predicted on Mars. The radiant areas of the northern and ecliptical
southern showers are the parts of a continuous area. Actually, we
initially identified a lower number of filaments when starting this
work based on the radiant positions. In the next stage, we noticed
that some particles, in the preliminary identified filaments, had
different characteristics than a main group and can be distinguished
as a separate group or more groups. We separated these groups as
extra filaments.

   Similarly to some showers predicted on Venus, also showers predicted
on Mars have a narrow, but extended-in-declination radiant area. This
property is remarkable in the case of the radiant area of NF1 (full
purple circles in Fig.~\ref{hammerMars}). A reason for the narrowness
of this radiant area is unknown. The perihelion distance close to the
orbit of Mars is not, probably, the reason, since there are other
filaments with such a large perihelion distance, but their radiant area
is not so narrow.

\subsection{On the common direction of radiants}

   When we compare Figs.~\ref{hammerVen} and \ref{hammerMars} with the
radiants of 21P's meteors on Venus and Mars, and Fig.~3a,c,e in Paper I
with the radiants of 21P's meteors on Earth, we can see that the
particles come to each of these planets from similar directions.

\begin{figure*}
\centerline{\includegraphics[width=5.6cm, angle=-90]{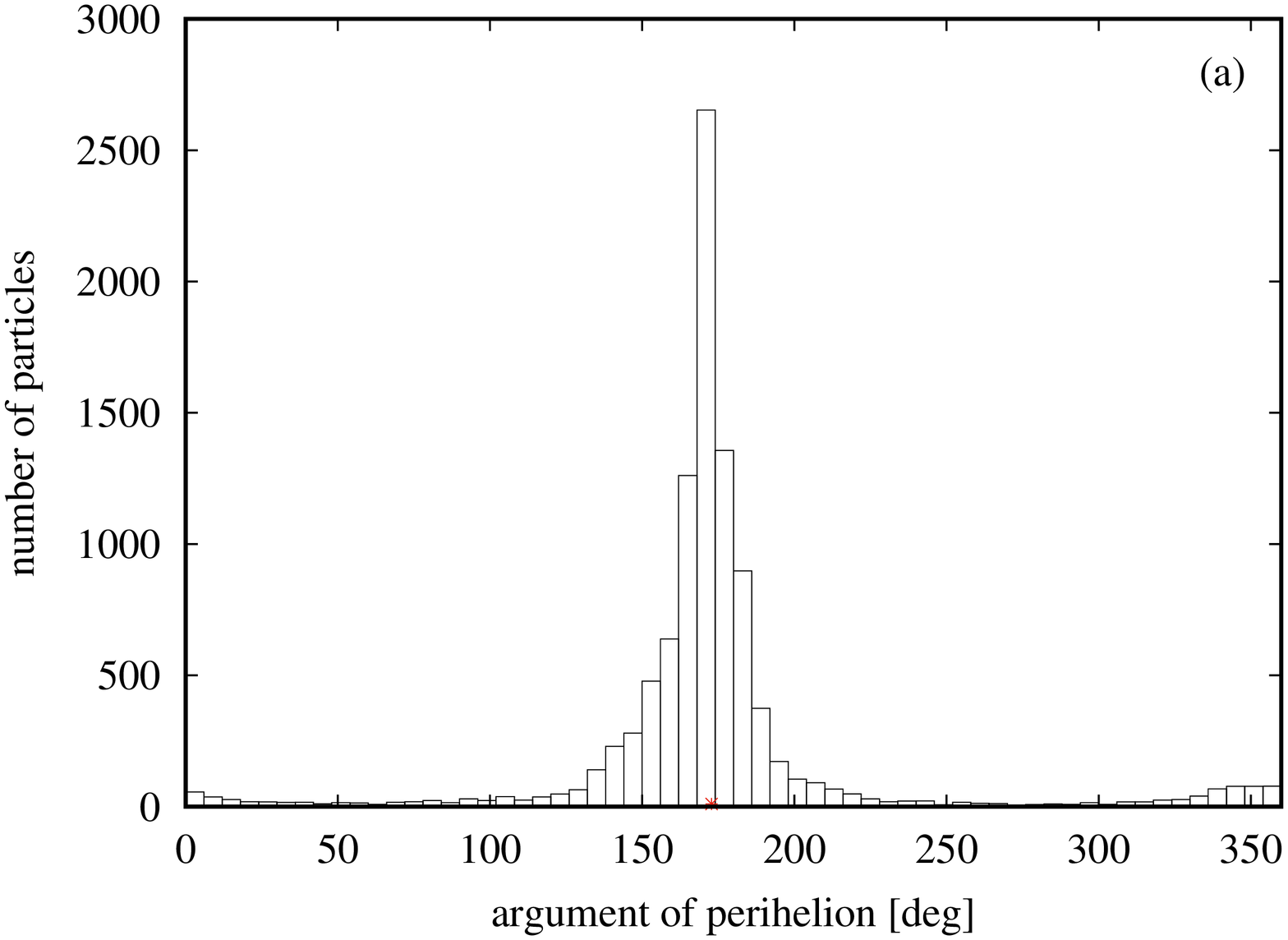}
            \includegraphics[width=5.6cm, angle=-90]{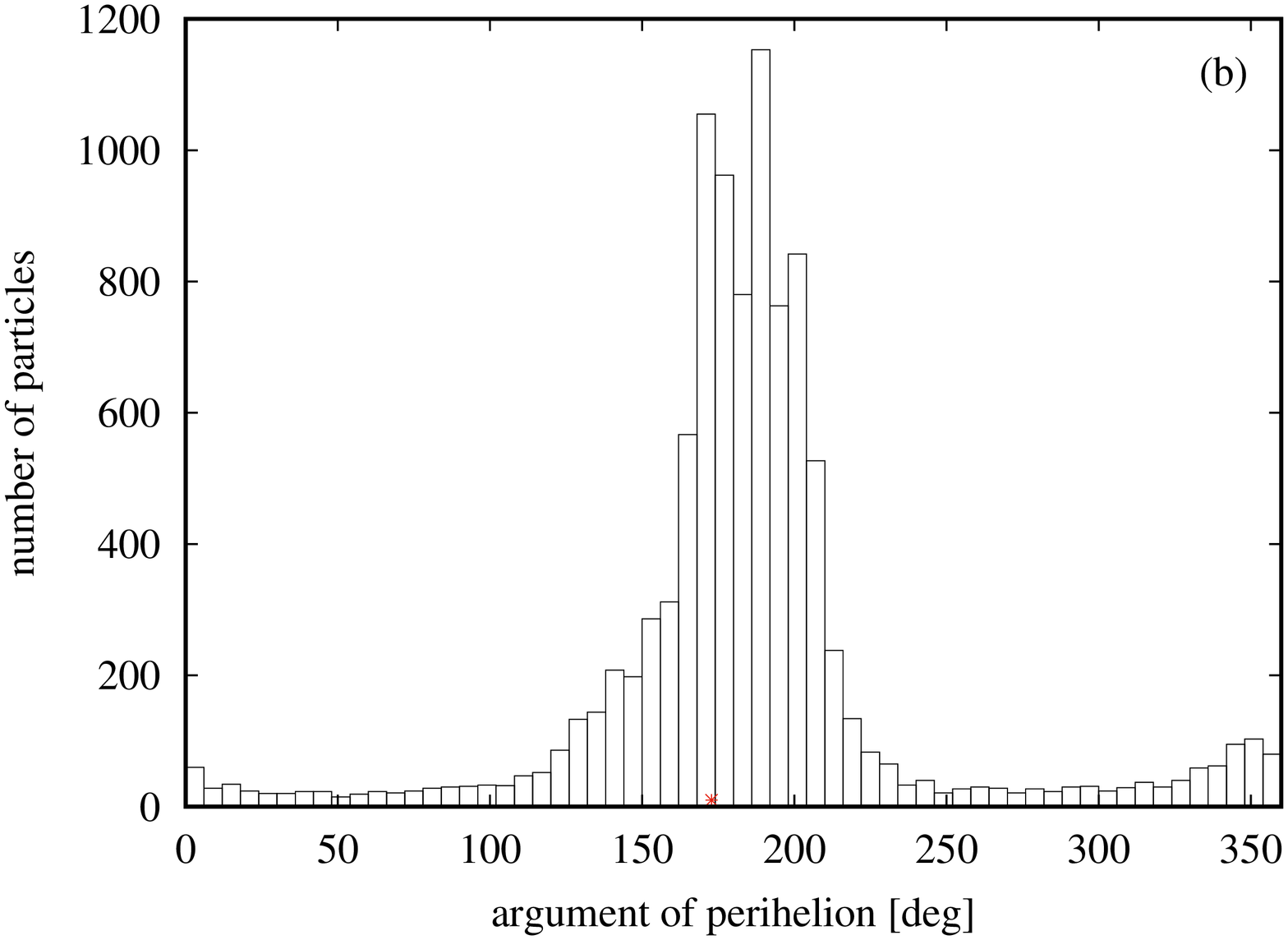}}
\centerline{\includegraphics[width=5.6cm, angle=-90]{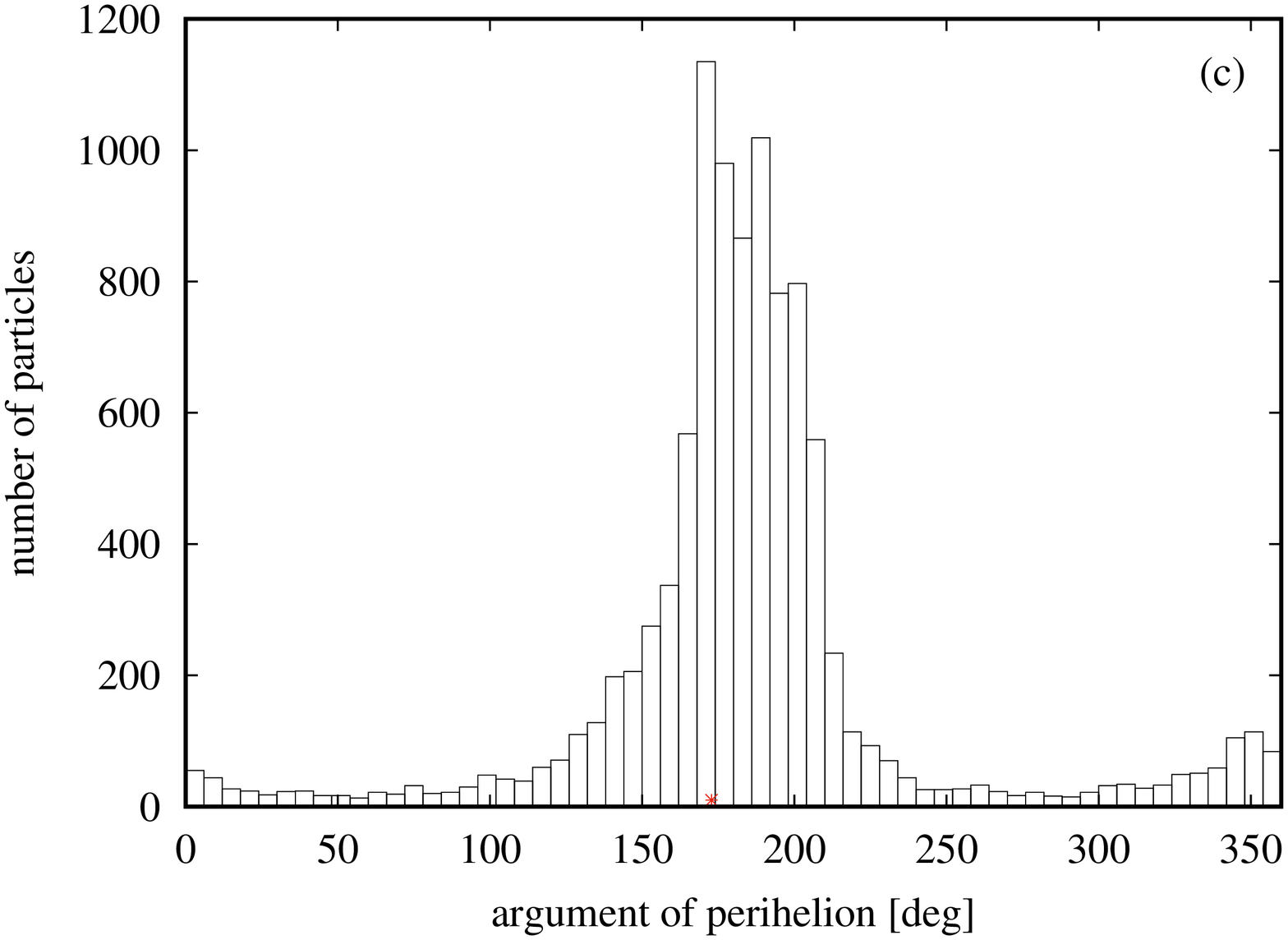}
            \includegraphics[width=5.6cm, angle=-90]{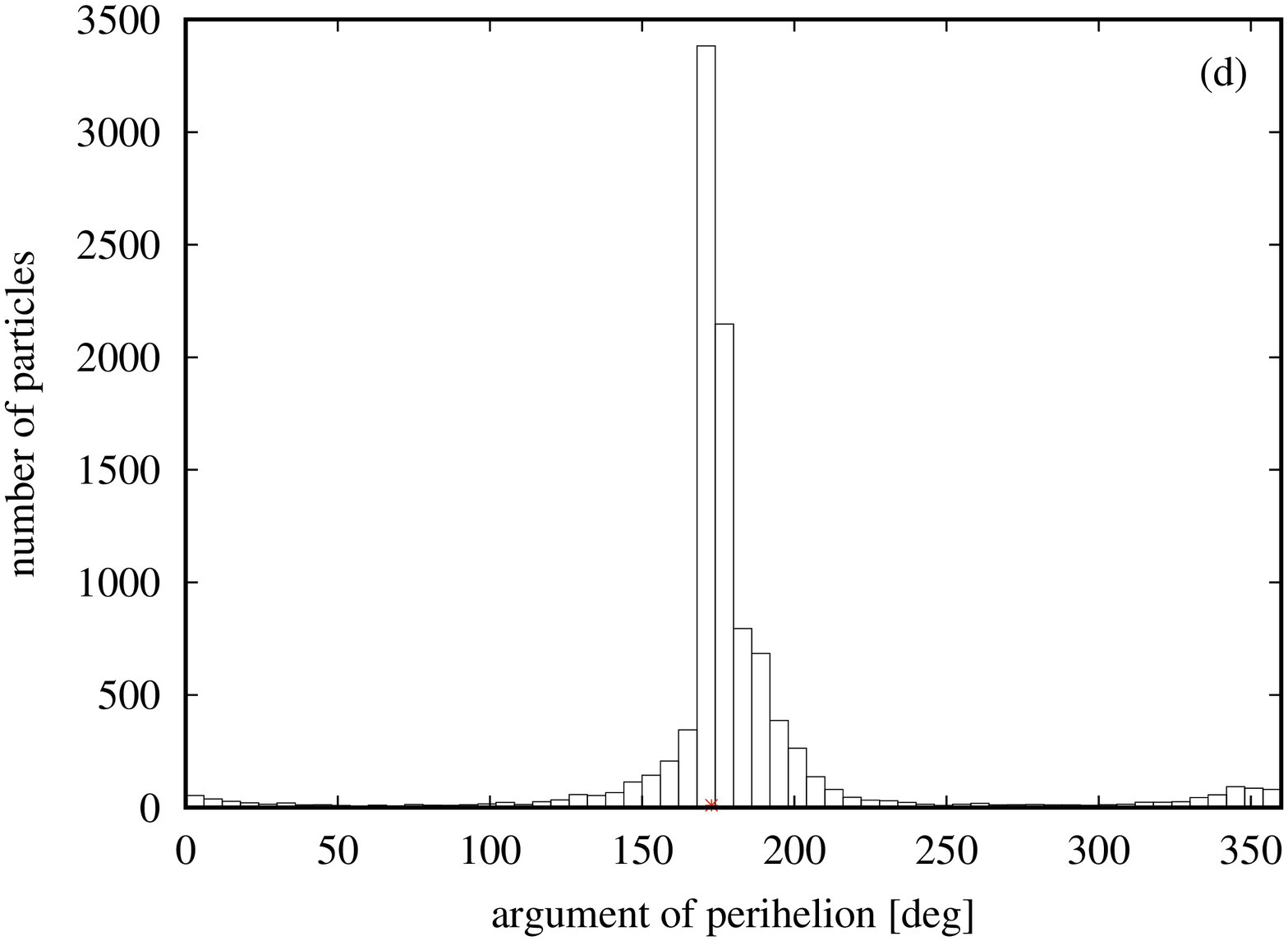}}
\centerline{\includegraphics[width=5.6cm, angle=-90]{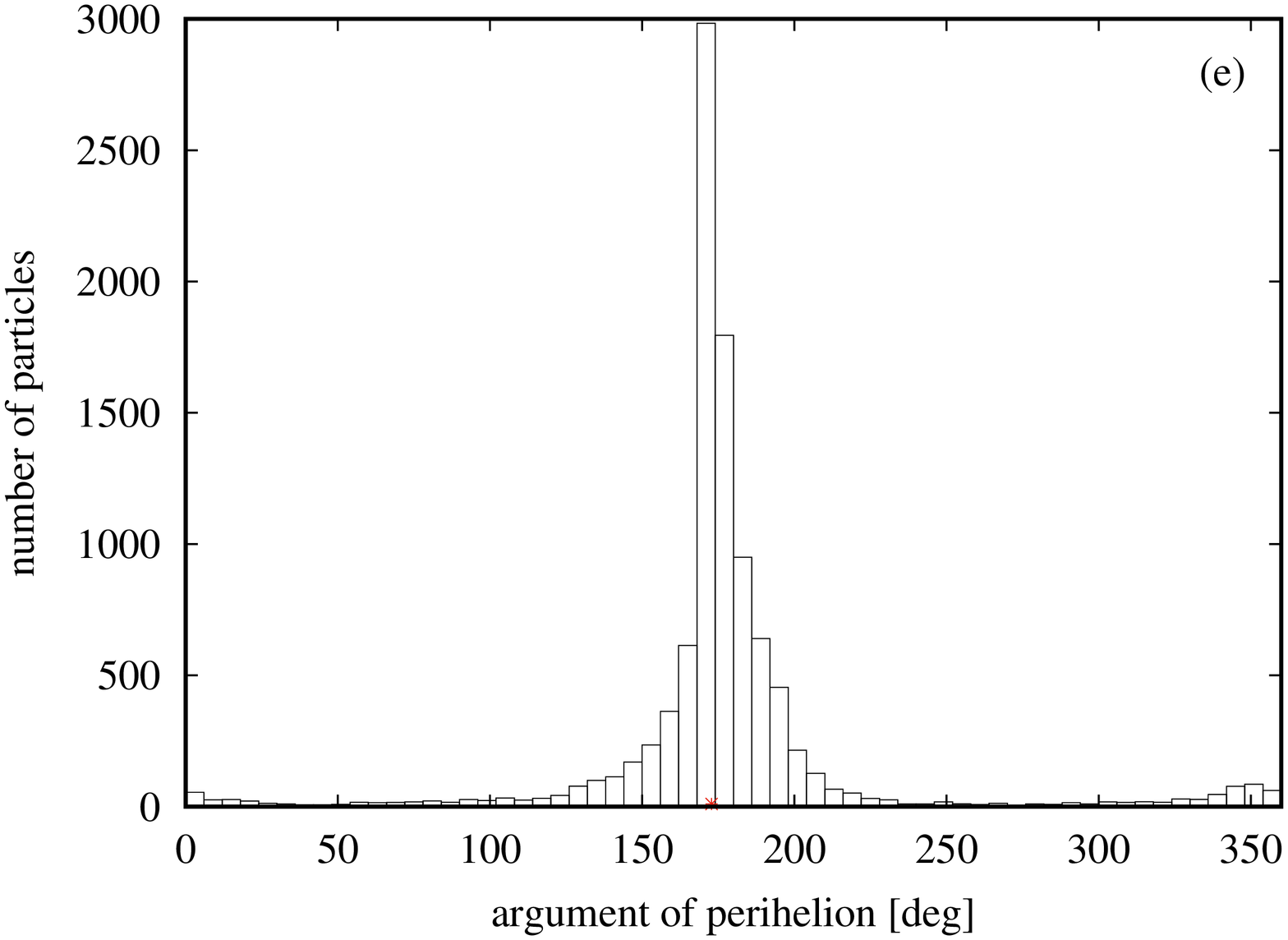}
            \includegraphics[width=5.6cm, angle=-90]{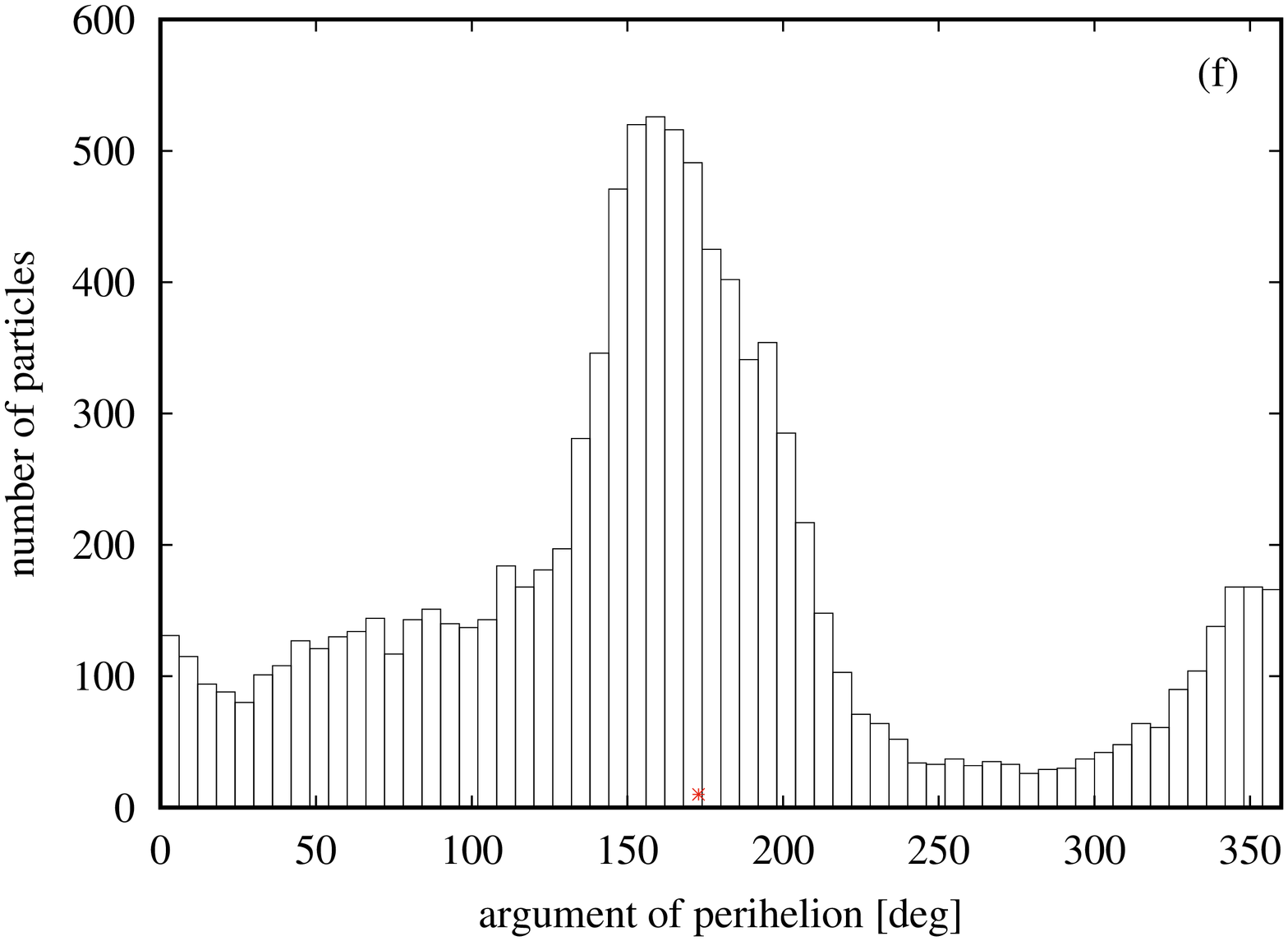}}
\caption{Distributions of the argument of perihelion, $\omega$, in the
models for $(t_{ev},~\beta) = (0.5\,$kyr, $10^{-11})$ (panel a),
$(1\,$kyr, $10^{-11})$ (b), $(1\,$kyr, $10^{-4})$ (c), $(1\,$kyr,
$0.003)$ (d), $(1\,$kyr, $0.005)$ (e), and $(2\,$kyr, $10^{-11})$ (f).
An asterisk at the horizontal coordinate axis indicates the value of
$\omega$ of the nominal orbit of parent comet.}
\label{FIGsom}
\end{figure*}

   If we analyze the orbital geometry of the particles in concerning
filaments, we can find some interesting features. The mean perihelia
of the filaments are situated at the orbit of given planet (cf.
$q \sim 0.72\,$au for the Venus' filaments, $q \sim 1.0\,$au for the
Earth's, and $q \sim 1.4\,$au for the Mars' filaments). The particles
can collide with a given planet only in an arc of their orbits close
to the perihelia because the argument of perihelion, $\omega$, of a
prevailing part of the whole stream is close to $180^{\circ}$
(Fig.~\ref{FIGsom}) or $0^{\circ}$ (the second, smaller peak in the
$\omega$-distribution). Since the distribution of longitude of ascending
node, $\Omega$, of the whole stream is also compact, peaked at the
values ranging from $\sim$$180^{\circ}$ to $\sim$$200^{\circ}$ (with
a small, second peak at $\sim$$10^{\circ}$ to $\sim$$40^{\circ}$), the
incoming direction of colliding particles is then almost unique.


   In Fig.~\ref{FIGsom}, we can see a high peak in the distribution
of $\omega$ in several selected models with the abundant filaments.
For $t_{ev} = 1\,$kyr, the peak is narrow as for a weak
($\beta = 10^{-11}$ or $\beta = 10^{-4}$) as for a relatively strong
($\beta = 0.003$ or $\beta = 0.005$) P-R effect. It is also narrow
in the model for $t_{ev} = 0.5\,$kyr and $\beta = 10^{-11}$ and its
width is not much larger either in the model for $t_{ev} = 2\,$kyr
and $\beta = 10^{-11}$. Despite the orbits of particles in the 21P'
stream are rapidly changing, the argument of perihelion seems to be
stable. The same can be observed for the longitude of ascending node
of a prevailing part of particles in stream. On the other-hand side,
the perihelia in the stream are predicted to be distributed throughout
almost whole region of terrestrial planets, from $\sim$$0.3$ to
$\sim$$2.4\,$au.

\section{Summary and conclusion}

   Using the resultant data of the modeling of meteoroid stream of comet
21P/Giacobini-Zinner, described in Paper I, we analyzed the close
approaches of the particles, representing the meteoroids, to the orbits
of Mercury, Venus, and Mars (the analysis concerning the Earth was done
in the past and results published in Paper I).

   We found several filaments of 21P' stream approaching the orbit of
each of these three planets. On the basis of mean characteristics of
these filaments, we predicted corresponding meteorite or meteor showers,
which could be detected on each planet. Specifically, we found 5 (4), 6
(3), and 11 (5) northern (southern) showers on Mercury, Venus, and Mars,
respectively. However, the radiants of particles of some predicted
showers were rather dispersed onto a relatively large area of sky. The
orbital elements were dispersed correspondingly. We classified these
showers as the less certain showers of category (ii). Of the total
number, 1 (2), 2 (1), and 2 (0) northern (southern) predicted showers
were the showers of category (ii) at Mercury, Venus, and Mars,
respectively.

   Taking into account also the result given in Paper I, which dealt
with the prediction of 21P' showers at the Earth, we found that,
except of few minor showers, the positions of radiant areas of the
showers predicted at Venus, Earth, and Mars are similar, i.e. the
meteoroids hit each planet incoming from a similar direction. We
explained this phenomenon by the orbital geometry of the 21P' stream:
the distributions of argument of perihelion and longitude of ascending
node have a dominant, high peak, which is conserved a relatively long
time despite a rapid change of orbits. The value of argument of
perihelion corresponding to the peak is close to $180^{\circ}$,
therefore the particles hit each planet at their perihelia.

   Our modeling also showed that a lot of particles predicted to hit
a terrestrial planet have a chaotic radiant distributions, therefore
they do not belong to any shower. Comet 21P significantly contributes
to the sporadic meteorites or meteors which can be detected on any
terrestrial planet.\\

{\bf Acknowledgements.} This work was supported, in part, by the VEGA
- the Slovak Grant Agency for Science, grant No. 2/0009/22, and by
the Slovak Research and Development Agency under the contract
No. APVV-19-0072.\\

\bibliographystyle{cas-model2-names}

\bibliography{tomko_neslusan}



\clearpage

\appendix
\section{Tabular information}


\begin{table*}
\caption{The mean planet-centric parameters of the showers associated
with the comet 21P/Giacobini-Zinner, which are predicted to be seen on
Mercury. In the individual columns, the following mean parameters are
given: $t_{ev}$ $-$ the evolutionary time; $\beta$ $-$ the parameter
characterizing the strength of non-gravitational force in respect to
the gravity of the Sun, $\lambda_{\odot}$ $-$ the mean solar longitude;
$\lambda_{\odot}^{min}$ and $\lambda_{\odot}^{max}$ are the minimum and
maximum solar longitudes of the shower; these values delimit the shower
activity; $\alpha$ and $\delta$ are the mean equatorial coordinates of
the planet-centric radiant; $V_{pc}$ is the mean planet-centric velocity;
$\gamma$ is the angular distance of mean radiant from the Sun in time
corresponding to $\lambda_{\odot}$; and $n$ is the number of particles
in the predicted shower (the total number of the particles in each model
was 10,000).}
\begin{tabular}{rccccccrr}
\toprule
$t_{ev}$ & $\beta$ & $\lambda_{\odot}$ & $\lambda_{min}$$-$$\lambda_{max}$
 & $\alpha$ & $\delta$ & $V_{pc}$ & $\gamma$ & $n$ \\
$[$kyr$]$ & $[$1$]$ & $[$deg$]$ & $[$deg$]$ & $[$deg$]$ & $[$deg$]$ &
 km$\,$s$^{-1}$ & $[$deg$]$ & \\
\midrule
\multicolumn{9}{c}{filament NF1} \\
   0.5 & $10^{-11}$ &  74.8 $\pm$  3.1 &  71.2$-$79.2  &   7.0 $\pm$  3.2 & $ 16.3 \pm 1.5$ & 22.8 $\pm$ 1.9 &  63. &   6 \\ 
   0.5 & $10^{-4}$  &  85.6 $\pm$  5.8 &  80.0$-$94.5  &  12.8 $\pm$  6.9 & $ 21.4 \pm 2.5$ & 19.2 $\pm$ 3.4 &  66. &   6 \\ 
   1   & $10^{-11}$ &  93.4 $\pm$  1.6 &  91.5$-$96.3  &  13.2 $\pm$  8.3 & $ 26.1 \pm 1.6$ & 16.6 $\pm$ 3.7 &  72. &  14 \\ 
   1   & $10^{-4}$  &  93.0 $\pm$  3.1 &  87.9$-$97.3  &  11.4 $\pm$  8.5 & $ 25.7 \pm 1.9$ & 16.1 $\pm$ 4.0 &  73. &  10 \\ 
   1   &  0.001     &  93.4 $\pm$  2.1 &  90.1$-$96.3  &  16.0 $\pm$  5.2 & $ 26.1 \pm 2.5$ & 18.0 $\pm$ 3.7 &  70. &   6 \\
\end{tabular}
\end{table*}

\addtocounter{table}{-1}
\begin{table*}
\caption{$-$ continuation.}
\begin{tabular}{rccccccrr}
\midrule
$t_{ev}$ & $\beta$ & $\lambda_{\odot}$ & $\lambda_{min}$$-$$\lambda_{max}$
 & $\alpha$ & $\delta$ & $V_{pc}$ & $\gamma$ & $n$ \\
$[$kyr$]$ & $[$1$]$ & $[$deg$]$ & $[$deg$]$ & $[$deg$]$ & $[$deg$]$ &
 km$\,$s$^{-1}$ & $[$deg$]$ & \\
\midrule
\multicolumn{9}{c}{filament SF1} \\
$[$0.5 & $10^{-11}$ & 309.3 $\pm$  7.0 & 297.6$-$322.9 & 300.1 $\pm$  7.2 & $-24.7 \pm 2.3$ & 31.1 $\pm$ 0.5 &  13. &  21$]$ \\ 
$[$0.5 & $10^{-4}$  & 306.5 $\pm$  4.3 & 301.3$-$315.1 & 297.8 $\pm$  3.9 & $-25.3 \pm 1.3$ & 30.8 $\pm$ 0.5 &  12. &  15$]$ \\ 
$[$0.5 &  0.001     & 312.0 $\pm$  7.1 & 300.1$-$324.5 & 302.3 $\pm$  6.9 & $-23.4 \pm 2.0$ & 31.4 $\pm$ 0.7 &  13. &  18$]$ \\ 
$[$0.5 &  0.003     & 302.5 $\pm$ 10.0 & 289.9$-$314.4 & 293.4 $\pm$ 10.8 & $-26.2 \pm 1.6$ & 30.8 $\pm$ 0.6 &  12. &   8$]$ \\ 
$[$1   & $10^{-11}$ & 321.3 $\pm$  6.4 & 307.8$-$329.3 & 312.9 $\pm$  6.6 & $-22.5 \pm 2.3$ & 31.0 $\pm$ 0.5 &  13. &  23$]$ \\
\bottomrule
\end{tabular}
\label{TAB1MERC}
\end{table*}
\vspace{-3 mm}


\begin{table*}
\caption{The mean orbital elements of the showers associated with the
comet 21P, which are predicted to be detected on Mercury. Quantities
$t_{ev}$ and $\beta$ are the same as in
Table~\ref{TAB1MERC}.}
\begin{tabular}{rcccccc}
\toprule
$t_{ev}$ & $\beta$ & $q$ & $e$ & $\omega$ & $\Omega$ & $i$ \\
$[$kyr$]$ & $[$1$]$ & $[$au$]$ & & $[$deg$]$ & $[$deg$]$ & $[$deg$]$ \\
\midrule
\multicolumn{7}{c}{filament NF1} \\
   0.5 & $10^{-11}$ & 0.352 $\pm$ 0.014 & 0.881 $\pm$ 0.004 & 355.6 $\pm$  1.6 &  13.5 $\pm$  5.1 &  18.7 $\pm$ 1.7 \\
   0.5 & $10^{-4}$  & 0.332 $\pm$ 0.016 & 0.886 $\pm$ 0.004 & 352.7 $\pm$  3.5 &  30.1 $\pm$  6.9 &  19.1 $\pm$ 2.3 \\
   1   & $10^{-11}$ & 0.329 $\pm$ 0.013 & 0.886 $\pm$ 0.004 & 349.8 $\pm$  3.7 &  41.7 $\pm$  5.6 &  20.6 $\pm$ 2.3 \\
   1   & $10^{-4}$  & 0.328 $\pm$ 0.013 & 0.887 $\pm$ 0.004 & 349.3 $\pm$  4.3 &  41.9 $\pm$  6.8 &  20.1 $\pm$ 2.0 \\
   1   &  0.001     & 0.327 $\pm$ 0.007 & 0.887 $\pm$ 0.002 & 351.4 $\pm$  3.9 &  40.4 $\pm$  6.0 &  21.3 $\pm$ 2.0 \\
\end{tabular}
\end{table*}

\addtocounter{table}{-1}
\begin{table*}
\caption{$-$ continuation.}
\begin{tabular}{rcccccc}
\midrule
$t_{ev}$ & $\beta$ & $q$ & $e$ & $\omega$ & $\Omega$ & $i$ \\
$[$kyr$]$ & $[$1$]$ & $[$au$]$ & & $[$deg$]$ & $[$deg$]$ & $[$deg$]$ \\
\midrule
\multicolumn{7}{c}{filament SF1} \\
$[$0.5 & $10^{-11}$ & 0.332 $\pm$ 0.017 & 0.888 $\pm$ 0.005 & 205.5 $\pm$ 24.3 & 169.1 $\pm$ 23.1 &   8.3 $\pm$ 4.5$]$ \\
$[$0.5 & $10^{-4}$  & 0.342 $\pm$ 0.018 & 0.885 $\pm$ 0.005 & 206.0 $\pm$ 24.1 & 167.2 $\pm$ 24.1 &   8.6 $\pm$ 3.7$]$ \\
$[$0.5 &  0.001     & 0.320 $\pm$ 0.024 & 0.891 $\pm$ 0.007 & 210.9 $\pm$ 27.1 & 165.3 $\pm$ 26.6 &   6.3 $\pm$ 2.9$]$ \\
$[$0.5 &  0.003     & 0.343 $\pm$ 0.023 & 0.886 $\pm$ 0.006 & 210.3 $\pm$ 15.7 & 159.0 $\pm$ 16.6 &   8.1 $\pm$ 3.3$]$ \\
$[$1   & $10^{-11}$ & 0.334 $\pm$ 0.018 & 0.887 $\pm$ 0.006 & 212.7 $\pm$ 23.6 & 174.0 $\pm$ 23.2 &   8.4 $\pm$ 4.2$]$ \\
\bottomrule
\end{tabular}
\label{TAB2MERC}
\end{table*}

\begin{table*}
\caption{The mean planet-centric parameters of the showers associated
with the comet 21P/Giacobini-Zinner, which are predicted to be seen on
Venus. The same denotation as in Table~\ref{TAB1MERC} is used.}
\begin{tabular}{lccccccrr}
\toprule
$t_{ev}$ & $\beta$ & $\lambda_{\odot}$ & $\lambda_{min}$$-$$\lambda_{max}$
 & $\alpha$ & $\delta$ & $V_{pc}$ & $\gamma$ & $n$ \\
$[$kyr$]$ & $[$1$]$ & $[$deg$]$ & $[$deg$]$ & $[$deg$]$ & $[$deg$]$ &
 km$\,$s$^{-1}$ & $[$deg$]$ & \\
\midrule
\multicolumn{9}{c}{filament NF1} \\
   0.5 & $10^{-11}$ & 193.7 $\pm$  2.6 & 182.8$-$199.4 & 275.7 $\pm$  5.6 & $ 59.8 \pm 2.0$  & 22.7 $\pm$ 0.7 &  91. &  67 \\ 
   0.5 & $10^{-4}$  & 194.0 $\pm$  2.8 & 183.1$-$201.0 & 274.6 $\pm$  5.1 & $ 59.7 \pm 2.1$  & 22.8 $\pm$ 0.8 &  91. &  66 \\ 
   0.5 &  0.001     & 193.5 $\pm$  1.3 & 187.2$-$197.2 & 277.0 $\pm$  2.4 & $ 59.8 \pm 1.3$  & 22.7 $\pm$ 0.4 &  92. &  52 \\ 
   1   & $10^{-4}$  & 198.6 $\pm$  3.0 & 193.1$-$204.1 & 276.6 $\pm$  7.3 & $ 63.0 \pm 2.2$  & 24.1 $\pm$ 1.0 &  92. &  35 \\ 
   1   &  0.001     & 197.7 $\pm$  4.1 & 191.0$-$205.4 & 277.9 $\pm$  9.9 & $ 61.7 \pm 2.7$  & 23.8 $\pm$ 1.0 &  92. &  32 \\
\end{tabular}
\end{table*}

\addtocounter{table}{-1}
\begin{table*}
\caption{$-$ continuation.}
\begin{tabular}{lccccccrr}
\midrule
$t_{ev}$ & $\beta$ & $\lambda_{\odot}$ & $\lambda_{min}$$-$$\lambda_{max}$
 & $\alpha$ & $\delta$ & $V_{pc}$ & $\gamma$ & $n$ \\
$[$kyr$]$ & $[$1$]$ & $[$deg$]$ & $[$deg$]$ & $[$deg$]$ & $[$deg$]$ &
 km$\,$s$^{-1}$ & $[$deg$]$ & \\
\midrule
\multicolumn{9}{c}{filament SF1} \\
$[$0.5 & $10^{-4}$  & 257.5 $\pm$  6.9 & 247.5$-$273.2 & 270.5 $\pm$  6.5 & $-29.9 \pm 3.1$  & 19.5 $\pm$ 1.0 &  14. &  13$]$ \\ 
$[$0.5 &  0.005     & 267.1 $\pm$  5.8 & 260.8$-$274.9 & 274.3 $\pm$  1.3 & $-28.9 \pm 1.6$  & 22.1 $\pm$ 2.2 &   9. &   8$]$ \\ 
$[$1   & $10^{-11}$ & 274.7 $\pm$ 28.9 & 143.6$-$304.2 & 291.4 $\pm$ 13.4 & $-29.1 \pm 3.3$  & 21.4 $\pm$ 2.2 &  16. &  26$]$ \\ 
$[$1   &  0.001     & 266.3 $\pm$ 32.5 & 132.6$-$287.4 & 283.2 $\pm$  8.7 & $-28.4 \pm 2.4$  & 21.7 $\pm$ 1.7 &  16. &  20$]$ \\ 
$[$1   &  0.003     & 256.9 $\pm$ 50.3 & 109.4$-$290.6 & 282.1 $\pm$  9.0 & $-28.8 \pm 3.3$  & 22.2 $\pm$ 1.4 &  24. &  28$]$ \\
\bottomrule
\end{tabular}
\label{TAB1VEN}
\end{table*}

\begin{table*}
\caption{The mean orbital elements of the showers associated with the
comet 21P, which are predicted to be detected on Venus. Quantities
$t_{ev}$ and $\beta$ are the same as in Table~\ref{TAB1MERC}.}
\begin{tabular}{lcccccr}
\toprule
$t_{ev}$ & $\beta$ & $q$ & $e$ & $\omega$ & $\Omega$ & $i$ \\
$[$kyr$]$ & $[$1$]$ & $[$au$]$ & & $[$deg$]$ & $[$deg$]$ & $[$deg$]$ \\
\midrule
\multicolumn{7}{c}{filament NF1} \\
  0.5  & $10^{-11}$ & 0.726 $\pm$ 0.010 & 0.772 $\pm$ 0.003 & 183.3 $\pm$  2.2 & 193.7 $\pm$  1.9 &  34.7 $\pm$ 1.4 \\
  0.5  & $10^{-4}$  & 0.723 $\pm$ 0.012 & 0.773 $\pm$ 0.004 & 182.9 $\pm$  2.0 & 193.8 $\pm$  2.1 &  34.8 $\pm$ 1.6 \\
  0.5  &  0.001     & 0.723 $\pm$ 0.012 & 0.773 $\pm$ 0.004 & 183.8 $\pm$  1.0 & 193.7 $\pm$  1.1 &  34.6 $\pm$ 0.9 \\
   1   & $10^{-11}$ & 0.725 $\pm$ 0.012 & 0.777 $\pm$ 0.009 & 186.4 $\pm$  5.7 & 192.4 $\pm$  7.5 &  31.0 $\pm$ 6.2 \\
   1   & $10^{-4}$  & 0.725 $\pm$ 0.012 & 0.768 $\pm$ 0.005 & 183.8 $\pm$  2.8 & 198.8 $\pm$  2.3 &  37.3 $\pm$ 2.0 \\
\end{tabular}
\end{table*}

\addtocounter{table}{-1}
\begin{table*}
\caption{$-$ continuation.}
\begin{tabular}{lcccccr}
\midrule
$t_{ev}$ & $\beta$ & $q$ & $e$ & $\omega$ & $\Omega$ & $i$ \\
$[$kyr$]$ & $[$1$]$ & $[$au$]$ & & $[$deg$]$ & $[$deg$]$ & $[$deg$]$ \\
\midrule
\multicolumn{7}{c}{filament SF1} \\
$[$0.5 & $10^{-4}$  & 0.699 $\pm$ 0.028 & 0.800 $\pm$ 0.006 & 344.3 $\pm$ 22.5 &  25.8 $\pm$ 24.3 &   7.7 $\pm$ 2.7$]$ \\
$[$0.5 &  0.005     & 0.639 $\pm$ 0.070 & 0.817 $\pm$ 0.014 & 309.6 $\pm$ 19.9 &  60.9 $\pm$ 18.2 &   4.3 $\pm$ 0.9$]$ \\
$[$1   & $10^{-11}$ & 0.644 $\pm$ 0.057 & 0.812 $\pm$ 0.013 & 315.8 $\pm$ 25.7 &  69.2 $\pm$ 24.1 &   6.4 $\pm$ 3.3$]$ \\
$[$1   &  0.001     & 0.648 $\pm$ 0.052 & 0.815 $\pm$ 0.011 & 318.9 $\pm$ 28.2 &  59.3 $\pm$ 26.3 &   5.1 $\pm$ 1.7$]$ \\
$[$1   &  0.003     & 0.630 $\pm$ 0.037 & 0.819 $\pm$ 0.008 & 321.3 $\pm$ 27.8 &  54.5 $\pm$ 29.7 &   5.6 $\pm$ 2.2$]$ \\
\bottomrule
\end{tabular}
\label{TAB2VEN}
\end{table*}

\begin{table*}
\caption{The mean planet-centric parameters of the showers associated
with the comet 21P/Giacobini-Zinner, which are predicted to be seen on
Mars. The same denotation as in Table~\ref{TAB1MERC} is used.}
\begin{tabular}{lccccccrr}
\toprule
$t_{ev}$ & $\beta$ & $\lambda_{\odot}$ & $\lambda_{min}$$-$$\lambda_{max}$
 & $\alpha$ & $\delta$ & $V_{pc}$ & $\gamma$ & $n$ \\
$[$kyr$]$ & $[$1$]$ & $[$deg$]$ & $[$deg$]$ & $[$deg$]$ & $[$deg$]$ &
 km$\,$s$^{-1}$ & $[$deg$]$ & \\
\midrule
\multicolumn{9}{c}{filament NF1} \\
 0.5     & $10^{-11}$ & 196.1 $\pm$  2.2 & 186.6$-$198.4 & 261.9 $\pm$  1.1 & $ 65.0 \pm 3.0$  & 23.1 $\pm$ 1.3 &  86. & 278 \\ 
 0.5     & $10^{-4}$  & 196.3 $\pm$  2.2 & 189.6$-$198.4 & 261.8 $\pm$  1.1 & $ 65.1 \pm 2.9$  & 23.2 $\pm$ 1.3 &  86. & 284 \\ 
 0.5     &  0.001     & 196.1 $\pm$  2.3 & 189.9$-$198.4 & 261.9 $\pm$  1.3 & $ 64.7 \pm 3.1$  & 23.0 $\pm$ 1.4 &  86. & 255 \\ 
 0.5     &  0.003     & 196.3 $\pm$  2.4 & 190.8$-$198.3 & 261.9 $\pm$  1.2 & $ 64.0 \pm 3.0$  & 22.8 $\pm$ 1.4 &  86. & 112 \\ 
 0.5     &  0.005     & 186.5 $\pm$  1.7 & 183.5$-$189.1 & 270.5 $\pm$  1.9 & $ 50.7 \pm 2.1$  & 17.3 $\pm$ 0.7 &  89. &  23 \\
\end{tabular}
\end{table*}

\addtocounter{table}{-1}
\begin{table*}
\caption{$-$ continuation.}
\begin{tabular}{lccccccrr}
\midrule
$t_{ev}$ & $\beta$ & $\lambda_{\odot}$ & $\lambda_{min}$$-$$\lambda_{max}$
 & $\alpha$ & $\delta$ & $V_{pc}$ & $\gamma$ & $n$ \\
$[$kyr$]$ & $[$1$]$ & $[$deg$]$ & $[$deg$]$ & $[$deg$]$ & $[$deg$]$ &
 km$\,$s$^{-1}$ & $[$deg$]$ & \\
\midrule
\multicolumn{9}{c}{filament NF4} \\
$[$0.5   & $10^{-11}$ &  98.2 $\pm$ 14.9 &  77.4$-$117.2 & 286.1 $\pm$ 12.4 & $-16.9 \pm 3.7$  & 28.7 $\pm$ 2.0 & 171. & 12$]$ \\ 
$[$0.5   & $10^{-4}$  &  82.1 $\pm$  6.7 &  72.6$-$ 91.7 & 272.2 $\pm$  4.9 & $-20.9 \pm 1.2$  & 30.2 $\pm$ 1.5 & 170. & 10$]$ \\ 
$[$0.5   &  0.001     &  87.9 $\pm$  6.9 &  78.4$-$ 99.9 & 276.8 $\pm$  5.0 & $-21.0 \pm 1.2$  & 29.3 $\pm$ 1.6 & 171. & 12$]$ \\ 
$[$0.5   &  0.003     &  70.8 $\pm$ 11.4 &  61.0$-$ 90.5 & 264.3 $\pm$  9.7 & $-21.9 \pm 0.6$  & 32.6 $\pm$ 2.4 & 166. &  7$]$ \\ 
$[$0.5   &  0.005     &  83.8 $\pm$ 20.5 &  62.0$-$115.8 & 269.4 $\pm$ 14.7 & $-19.6 \pm 2.9$  & 28.0 $\pm$ 3.6 & 173. &  5$]$ \\
\end{tabular}
\end{table*}

\addtocounter{table}{-1}
\begin{table*}
\caption{$-$ continuation.}
\begin{tabular}{lccccccrr}
\midrule
$t_{ev}$ & $\beta$ & $\lambda_{\odot}$ & $\lambda_{min}$$-$$\lambda_{max}$
 & $\alpha$ & $\delta$ & $V_{pc}$ & $\gamma$ & $n$ \\
$[$kyr$]$ & $[$1$]$ & $[$deg$]$ & $[$deg$]$ & $[$deg$]$ & $[$deg$]$ &
 km$\,$s$^{-1}$ & $[$deg$]$ & \\
\midrule
\multicolumn{9}{c}{filament SF1} \\
   0.5  & $10^{-4}$  & 230.5 $\pm$ 20.3 & 204.2$-$257.2 & 257.3 $\pm$  9.5 & $-29.5 \pm 1.4$  & 15.7 $\pm$ 2.7 &  29. &  8 \\ 
$[$0.5  &  0.003     & 248.9 $\pm$  8.9 & 239.1$-$264.6 & 264.9 $\pm$  5.4 & $-28.5 \pm 2.1$  & 18.5 $\pm$ 1.9 &  17. &  6$]$ \\ 
$[$1    & $10^{-11}$ & 247.6 $\pm$ 27.9 & 195.6$-$280.1 & 274.3 $\pm$ 13.8 & $-28.1 \pm 1.8$  & 17.5 $\pm$ 5.1 &  27. & 18$]$ \\ 
$[$1    &  0.003     & 243.5 $\pm$ 16.7 & 215.9$-$260.1 & 271.7 $\pm$  4.8 & $-30.3 \pm 2.5$  & 16.4 $\pm$ 3.6 &  29. & 10$]$ \\ 
$[$1    &  0.005     & 245.2 $\pm$ 29.2 & 159.8$-$262.7 & 272.0 $\pm$ 13.3 & $-29.5 \pm 3.2$  & 18.7 $\pm$ 2.3 &  27. & 11$]$ \\
\bottomrule
\end{tabular}
\label{TAB1MARS}
\end{table*}

\begin{table*}
\caption{The mean orbital elements of the showers associated with the
comet 21P, which are predicted to be detected on Mars. Quantities
$t_{ev}$ and $\beta$ are the same as in Table~\ref{TAB1MERC}.}
\begin{tabular}{lcccccc}
\toprule
$t_{ev}$ & $\beta$ & $q$ & $e$ & $\omega$ & $\Omega$ & $i$ \\
$[$kyr$]$ & $[$1$]$ & $[$au$]$ & & $[$deg$]$ & $[$deg$]$ & $[$deg$]$ \\
\midrule
\multicolumn{7}{c}{filament NF1} \\
 0.5     & $10^{-11}$ & 1.401 $\pm$ 0.014 & 0.615 $\pm$ 0.013 & 169.9 $\pm$  3.8 & 199.2 $\pm$  1.2 &  37.7 $\pm$ 2.6 \\
 0.5     & $10^{-4}$  & 1.402 $\pm$ 0.015 & 0.614 $\pm$ 0.012 & 170.0 $\pm$  3.7 & 199.4 $\pm$  1.1 &  37.9 $\pm$ 2.5 \\
 0.5     &  0.001     & 1.399 $\pm$ 0.014 & 0.617 $\pm$ 0.013 & 169.4 $\pm$  3.8 & 199.3 $\pm$  1.1 &  37.5 $\pm$ 2.8 \\
 0.5     &  0.003     & 1.394 $\pm$ 0.012 & 0.621 $\pm$ 0.012 & 168.5 $\pm$  3.7 & 199.8 $\pm$  1.0 &  37.0 $\pm$ 2.7 \\
 0.5     &  0.005     & 1.373 $\pm$ 0.008 & 0.667 $\pm$ 0.004 & 161.6 $\pm$  1.0 & 194.8 $\pm$  1.4 &  26.1 $\pm$ 1.5 \\
\end{tabular}
\end{table*}

\addtocounter{table}{-1}
\begin{table*}
\caption{$-$ continuation.}
\begin{tabular}{lcccccc}
\midrule
$t_{ev}$ & $\beta$ & $q$ & $e$ & $\omega$ & $\Omega$ & $i$ \\
$[$kyr$]$ & $[$1$]$ & $[$au$]$ & & $[$deg$]$ & $[$deg$]$ & $[$deg$]$ \\
\midrule
\multicolumn{7}{c}{filament NF4} \\
$[$0.5 & $10^{-11}$ & 0.415 $\pm$ 0.070 & 0.866 $\pm$ 0.018 & 307.1 $\pm$ 19.1 &  77.4 $\pm$ 21.9 &   6.2 $\pm$ 2.8$]$ \\
$[$0.5 & $10^{-4}$  & 0.362 $\pm$ 0.051 & 0.881 $\pm$ 0.014 & 301.2 $\pm$ 18.5 &  73.0 $\pm$ 19.4 &   3.2 $\pm$ 1.3$]$ \\
$[$0.5 &  0.001     & 0.389 $\pm$ 0.055 & 0.873 $\pm$ 0.014 & 312.8 $\pm$ 21.6 &  64.1 $\pm$ 22.5 &   3.0 $\pm$ 1.0$]$ \\
$[$0.5 &  0.003     & 0.283 $\pm$ 0.073 & 0.903 $\pm$ 0.021 & 326.0 $\pm$ 22.1 &  46.3 $\pm$ 18.1 &   2.1 $\pm$ 1.0$]$ \\
$[$0.5 &  0.005     & 0.437 $\pm$ 0.129 & 0.861 $\pm$ 0.029 & 297.0 $\pm$ 26.3 &  70.2 $\pm$ 27.9 &   3.4 $\pm$ 1.6$]$ \\
\end{tabular}
\end{table*}

\addtocounter{table}{-1}
\begin{table*}
\caption{$-$ continuation.}
\begin{tabular}{lcccccc}
\midrule
$t_{ev}$ & $\beta$ & $q$ & $e$ & $\omega$ & $\Omega$ & $i$ \\
$[$kyr$]$ & $[$1$]$ & $[$au$]$ & & $[$deg$]$ & $[$deg$]$ & $[$deg$]$ \\
\midrule
\multicolumn{7}{c}{filament SF1} \\
 0.5   & $10^{-4}$  & 0.855 $\pm$ 0.081 & 0.771 $\pm$ 0.015 & 288.3 $\pm$  8.6 &  76.6 $\pm$ 15.1 &   3.4 $\pm$ 0.7 \\
$[$0.5 &  0.003     & 0.768 $\pm$ 0.064 & 0.790 $\pm$ 0.013 & 294.6 $\pm$ 12.7 &  74.5 $\pm$ 13.6 &   2.9 $\pm$ 1.4$]$ \\
$[$1   & $10^{-11}$ & 0.786 $\pm$ 0.154 & 0.787 $\pm$ 0.029 & 299.9 $\pm$ 14.3 &  74.6 $\pm$ 13.6 &   3.0 $\pm$ 1.5$]$ \\
$[$1   &  0.003     & 0.829 $\pm$ 0.099 & 0.781 $\pm$ 0.019 & 305.8 $\pm$ 24.9 &  69.7 $\pm$ 23.3 &   3.4 $\pm$ 1.4$]$ \\
$[$1   &  0.005     & 0.764 $\pm$ 0.077 & 0.795 $\pm$ 0.013 & 309.4 $\pm$ 29.9 &  62.8 $\pm$ 28.3 &   3.6 $\pm$ 1.5$]$ \\
\bottomrule
\end{tabular}
\label{TAB2MARS}
\end{table*}


\end{document}